\documentclass{ptapap}
\usepackage{color}

\usepackage{soul}

\author{Paweł Zieliński}[OAUW]
\author{Jan Jan{\'i}k}[MUNI]
\author{Ralph Neuh{\"a}user}[JENA]
\author{Markus Mugrauer}[JENA]
\author{Zoltan Garai}[AISAS]
\author{Theodor Pribulla}[AISAS]
\author{Marek Dróżdż}[UP]
\author{Waldemar Ogłoza}[UP]
\author{YETI Team}

\affil[OAUW]{Warsaw University Astronomical Observatory, Al. Ujazdowskie 4, 00--478 Warszawa, Poland}
\affil[MUNI]{Department of Theoretical Physics and Astrophysics, Faculty of Science, Masaryk University, Kotlářská 2, 611 37 Brno, Czech Republic}
\affil[JENA]{Astrophysical Institute and University Observatory, Friedrich Schiller University, Schillerg{\"a}ßchen 2-3, 07745 Jena, Germany}
\affil[AISAS]{Astronomical Institute, Slovak Academy of Sciences, 059 60 Tatransk{\'a}~Lomnica, Slovak Republic}
\affil[UP]{Mt.~Suhora Observatory, Pedagogical University, ul. Podchorążych 2, 30--084 Kraków, Poland}

\title{Search for young transiting exoplanets within YETI project}

\begin{document}

\maketitle

\begin{abstract}
The Young Exoplanet Transit Initiative (YETI) is a project focused on the photometric monitoring of stellar open clusters in order to find new young transiting exoplanets, eclipsing binaries and study other variability phenomena. Here, we present the status of the initiative and plans for future photometric campaigns of three open clusters younger than 50 Myr: NGC~869, NGC~884 and IC~4665, by using the world-wide one meter-class telescope network. Based on the experience gained by several astronomical observatories included in this network, dedicated numerical algorithms and recent results obtained during the first observing campaigns, we expect to confirm several young transiting objects: low-mass stars, brown dwarfs and exoplanets. The photometric precision given for a typical telescope used in this project, allows for transit detection of Jupiter-size planets at close-in orbits with periods up to {$\sim30$} days and also hundreds of new various variable stars.

\end{abstract}

\section{Introduction}
The primary goal of YETI is to search for transiting extrasolar planets and eclipsing binaries younger than 100 Myr and, therefore, to deliver a number of the youngest exoplanetary systems presently known. A detailed study of several carefully selected eclipsing binaries detected in open clusters with known ages and distances is necessary in order to determine reliable orbital and physical parameters of companions (exoplanets, brown-dwarfs, or low-mass stars of late spectral types). Eclipsing binaries detected in the low-mass regime can deliver valuable observational constraints on the mass-radius relation, which remains scarce in the domain of pre-main sequence stars and brown dwarfs. As a byproduct, we investigate also other types of variabilities in the monitored stars on different time-scales, including rotating and pulsating stars, irregular variables, flares, T~Tauri stars, etc., which are extremely interesting from the astrophysical point of view.

\begin{figure}
\centering
\includegraphics[width=\textwidth]{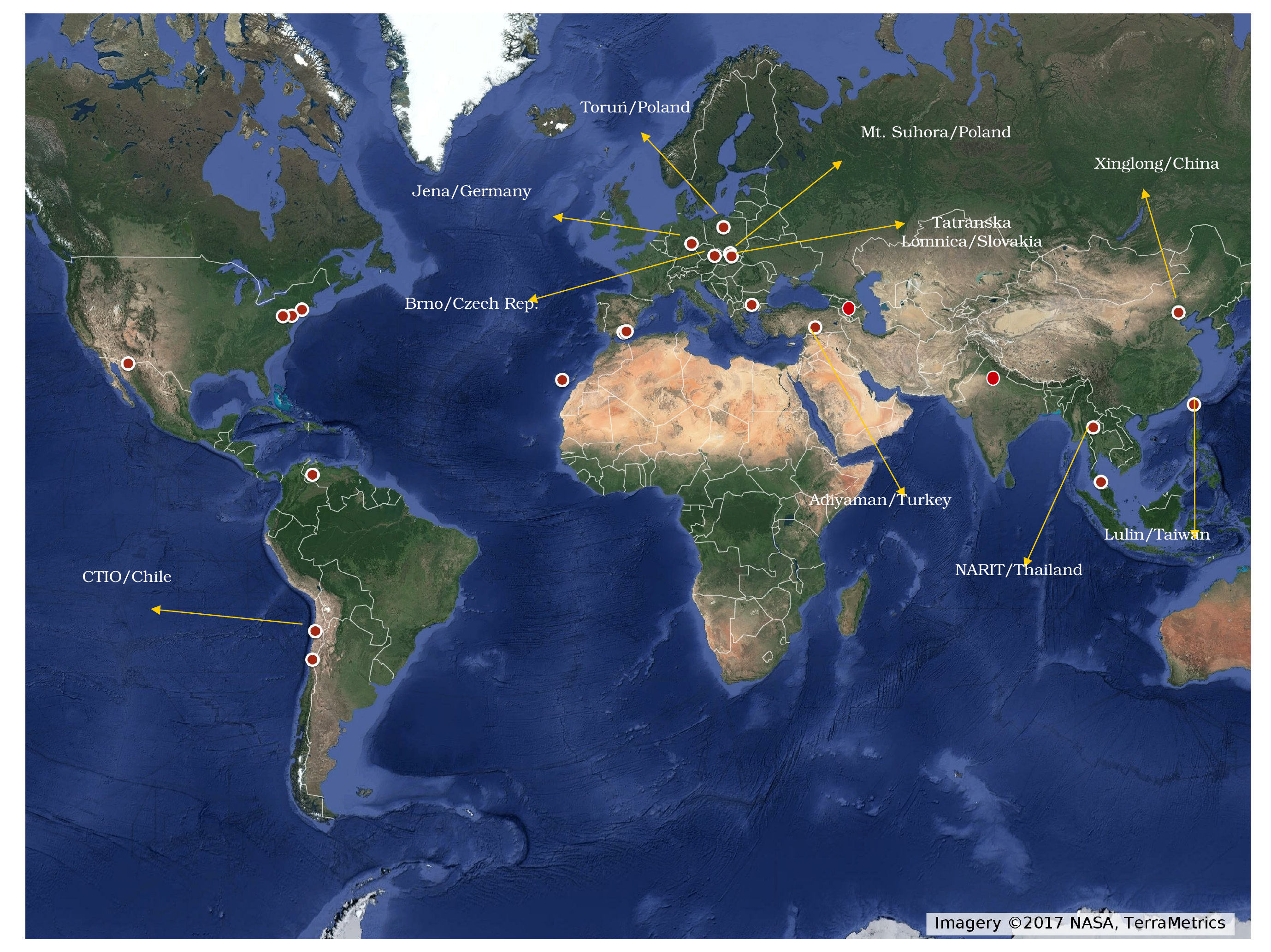}
\caption{The map of all YETI observatories (red dots) since 2010. The sites active in recent studies of clusters: NGC~869, NGC~884 and IC~4665 are signed.}
\label{fig:yetinet}
\end{figure}

The long-term motivation of YETI is to test star and planet formation scenarios by gathering robust observational indications on the early evolution of low-mass stellar, substellar and planetary objects. Despite the fact that we know more than 3700 exoplanets\footnote{after http://exoplanet.eu (15.12.2017)} today, there is a lack of confirmed transiting exoplanets around stars younger than 100 Myr (there are only four such exoplanets, where one of them -- CVSO~30b was confirmed within YETI by \citealt{Raetz2016}). The vast majority of clusters surveyed for transits are several Gyr old and this is one of the main reasons why we cannot answer the question whether giant planets form by gravitational contraction or by core accretion. In addition, among low-mass pre-main sequence stars, there are only very few eclipsing spectroscopic binary stars known, so there is a lack of precise parameter  determinations. None of the known models can give correct masses for low-mass pre-main sequence stars -- the models are still not correct for ages below 10 Myr. To improve this situation, the YETI network (Fig.~\ref{fig:yetinet}) was established by searching for transiting planets and eclipsing binaries in young clusters.

\section{Target clusters and observations}
\begin{figure}
\centering
\includegraphics[width=0.9\textwidth]{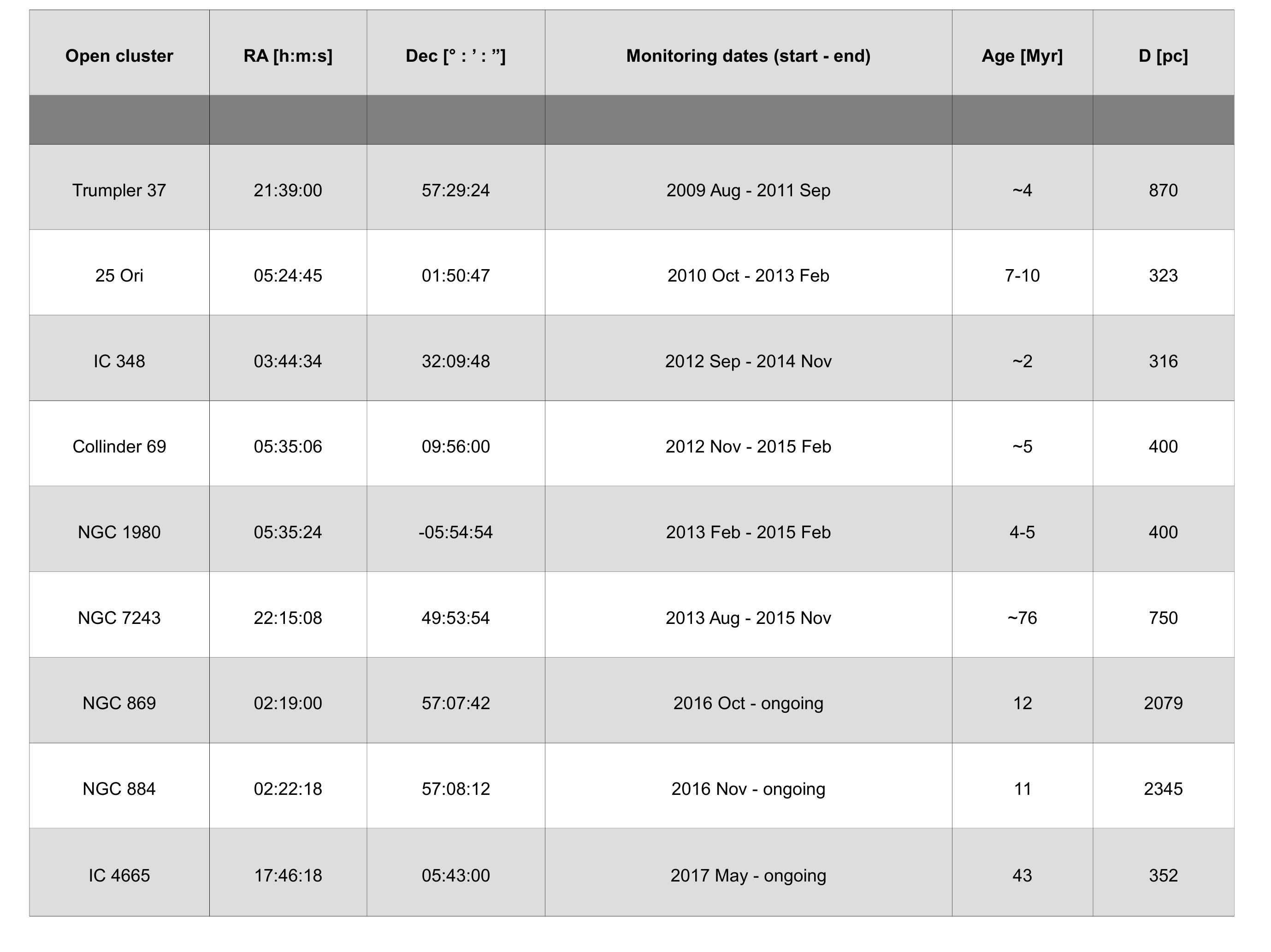}
\caption{The complete list of all YETI target clusters. The photometric monitoring for six of them were finished already, but the last three clusters are currently investigated.}
\label{fig:targets}
\end{figure}

The open clusters presented in Fig.~\ref{fig:targets} were selected for the photometric monitoring within the YETI network based on several assumptions mainly related to the young age of the cluster (younger than 100~Myr which is the pre-main sequence time-scale of the lowest mass stars), the intermediate distance, the size and location of the cluster on the sky.

Observational strategy within the YETI campaign is based on photometric monitoring of target clusters by using small class telescopes, roughly three times per year per cluster where one run is typically 14 days long. During the campaign, the CCD observations are conducted only with the Bessell R-band filter, by applying alternating short and long exposures to accommodate bright and faint stars in the cluster. Based on the modern methods of data reduction and analysis of variable stars we are able to detect of any transit with a depth of at least 5~mmag rms down to $R = 14.5$~mag, and 50~mmag rms down to $R = 16.5$~mag stars. For targets with periodic transit-like signals the follow-up observations are carried out (i.e., high-quality photometry to improve transit light curve, low- and high-resolution spectroscopy to ensure that the star is young and confirm the substellar nature of the companion by RV (radial velocity) variation analysis, adaptive optics IR (infrared) imaging to exclude other bright eclipsing stars in the larger PSF (point spread function).

\section{Results overview}
\begin{itemize}
\item {\bf Transiting planet candidates} \\{
Several transiting candidates were already identified in the YETI data. For the first candidate \citep{Errmann2014} all follow-ups are already done, while the other candidates needs further investigations \citep{Garai2016, Fritzewski2016}. High-resolution imaging and RV spectroscopy for the first candidate yields a companion mass consistent with that of a M5 to M6 dwarf, hence it is a false positive. The most promising and extremely unique planetary candidate so far is the system found around the star CVSO~30. This is the youngest known exoplanet system with both a close-in transiting \citep{Raetz2016} and a wide directly imaged \citep{Schmidt2016} planet candidates orbiting their host star.
}
\item {\bf Eclipsing binaries} \\{
We found hundreds of new eclipsing binaries in the YETI photometric measurements \citep{Errmann2013, Garai2016, Fritzewski2016}. Only for Trumpler 37 we found more than 50 new eclipsing binaries. The study of binary stars clearly shows the importance of using the telescope network in order to fully complete the phase-coverage of brightness changes \citep{Neuhauser2011}.
}
\item {\bf Rotation periods of young stars} \\{
We can also study the angular momentum evolution of low-mass pre-main sequence stars during their contraction toward the main sequence. The rotation period distribution of the 
T~Tauri stars in 25~Ori \citep{Karim2016} and IC~348 \citep{Fritzewski2016} shown that these stars typically rotate at a small fraction of their breakup velocity, despite the significant contraction in stellar radius, what is the subject of the discussion. 
}
\item {\bf Other variability phenomena} \\{
Except of different types of variable stars (rotating and pulsating stars with $1$~h $< P < 300$~ d, flaring stars, T~Tauri type stars, irregular variables) discovered in the YETI data, we are able to study already known variables. For example, GM~Cep has been known to be an abrupt variable and to have a circumstellar disk with a very active accretion. Our observations reveal $\sim$ 1~mag brightness drops with period of $\sim$2~yr. These drops might be caused by the obscuration of the central star by an orbiting clump of dust in the protoplanetary disk around the star, as described in \cite{Chen2012}.
}
\end{itemize}

\section{Future outlook}

\begin{figure}
\centering
\includegraphics[width=0.8\textwidth]{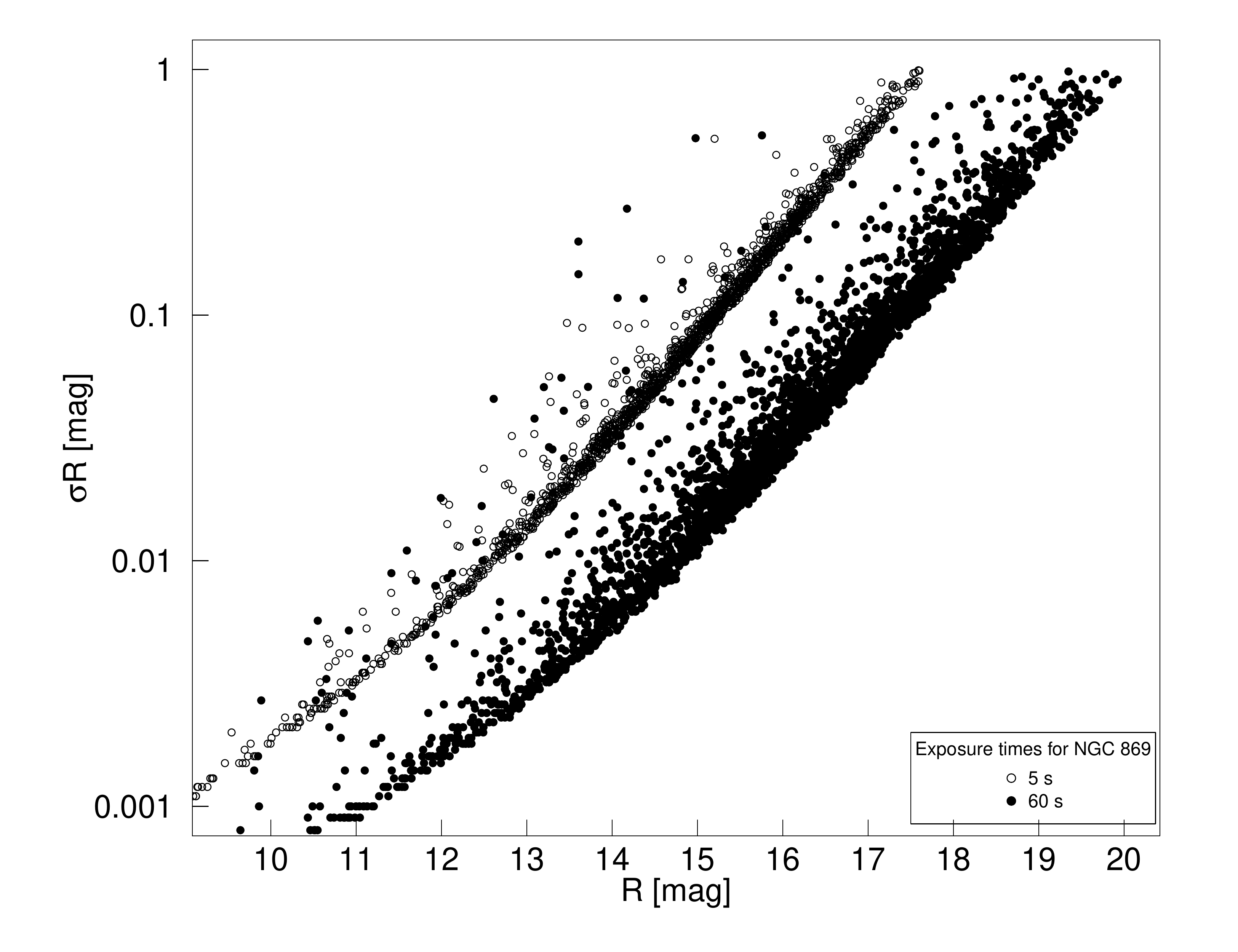}
\caption{Example diagram presenting the photometric precision for stars observed in the field-of-view of the cluster NGC~869. Data taken by the 0.6~m telescope at Mt. Suhora Observatory during one night. Open circles show short exposure and filled circles -- long exposure times. The brightness of stars in R band is presented on the x-axis, while on the y-axis, the precision in photometric measurements in R band is shown. Note the logarithmic scale on the y-axis.}
\label{fig:photprecision}
\end{figure}

The small telescopes that form the YETI network enable us to detect transits of extrasolar planets. The photometric precision acquired within YETI project is sufficient for the detection of Jupiter-size planets around G and K stars in the cluster fields. On the example diagram (Fig.~\ref{fig:photprecision}) we show that for the brightest stars a standard deviation of less than 5~mmag in the light curve is reached in a single night. For 1200 stars brighter than $R=16$~mag and observed in the FoV of NGC 869 it is better than 30~mmag. Young clusters including Trumpler~37, 25~Ori, IC~348, Collinder~69, NGC~1980 and NGC~7243 were searched for transits by YETI for several weeks and the monitoring is finished at this moment. For our first candidates we have completed all follow-up observations to reject false positives and measure the mass of the companion. For the other cluster candidates follow-up observations are in progress and the results should be published soon. Recent observations of NGC~869, NGC~884 and IC~4665 yield to detections of several promising candidates for transiting planets. However, period determinations of these transits will be possible after the reduction of all data. We expect to detect further candidates among the cluster members from our ongoing observations as well as from further data reduction and improved transit search algorithms. Taking into account an optimistic case, we will be able to increase the number of known young transiting planets by a factor of up to three.

\acknowledgements{This research relies on work of many people involved in the YETI network. We especially thank to all involved YETI observers from Jena, Tatransk{\'a} Lomnica, Toruń, Mt.Suhora, Brno, Adiyaman, NARIT, CTIO, Xinglong and Lulin observatories. This work was supported by the VEGA grant of the Slovak Academy of Sciences No.~2/0031/18 and by the realization of the Project ITMS No.~26220120029, based on the Supporting
Operational Research and Development Program financed from the European Regional Development Fund. ZG thanks the support from the Slovak Central Observatory Hurbanovo.}

\bibliographystyle{ptapap}
\bibliography{zielinski_literatura}

\begin{thebibliography}{9}
\providecommand{\natexlab}[1]{#1}
\providecommand{\url}[1]{\texttt{#1}}
\providecommand{\urlprefix}{URL }
\providecommand{\eprint}[2][]{\url{#2}}

\bibitem[{{Chen} et~al.(2012)}]{Chen2012}
{Chen}, W.~P., et~al., \emph{{A Possible Detection of Occultation by a
  Proto-planetary Clump in GM Cephei}}, \emph{\apj} \textbf{751}, 118 (2012),
  \eprint{1203.5271}

\bibitem[{{Errmann} et~al.(2013)}]{Errmann2013}
{Errmann}, R., et~al., \emph{{YETI - search for young transiting planets}}, in
  European Physical Journal Web of Conferences, \emph{European Physical Journal
  Web of Conferences}, volume~47, 03004 (2013)

\bibitem[{{Errmann} et~al.(2014)}]{Errmann2014}
{Errmann}, R., et~al., \emph{{Investigation of a transiting planet candidate in
  Trumpler 37: An astrophysical false positive eclipsing spectroscopic binary
  star}}, \emph{Astronomische Nachrichten} \textbf{335}, 345 (2014),
  \eprint{1403.6020}

\bibitem[{{Fritzewski} et~al.(2016)}]{Fritzewski2016}
{Fritzewski}, D.~J., et~al., \emph{{Long-term photometry of IC 348 with the
  Young Exoplanet Transit Initiative network}}, \emph{\mnras} \textbf{462},
  2396 (2016)

\bibitem[{{Garai} et~al.(2016)}]{Garai2016}
{Garai}, Z., et~al., \emph{{Search for transiting exoplanets and variable stars
  in the open cluster NGC 7243}}, \emph{Astronomische Nachrichten}
  \textbf{337}, 261 (2016), \eprint{1601.04562}

\bibitem[{{Karim} et~al.(2016)}]{Karim2016}
{Karim}, M.~T., et~al., \emph{{The Rotation Period Distributions of 4-10 Myr T
  Tauri Stars in Orion OB1: New Constraints on Pre-main-sequence Angular
  Momentum Evolution}}, \emph{\aj} \textbf{152}, 198 (2016),
  \eprint{1605.04333}

\bibitem[{{Neuh{\"a}user} et~al.(2011)}]{Neuhauser2011}
{Neuh{\"a}user}, R., et~al., \emph{{The Young Exoplanet Transit Initiative
  (YETI)}}, \emph{Astronomische Nachrichten} \textbf{332}, 547 (2011),
  \eprint{1106.4244}

\bibitem[{{Raetz} et~al.(2016)}]{Raetz2016}
{Raetz}, S., et~al., \emph{{YETI observations of the young transiting planet
  candidate CVSO 30 b}}, \emph{\mnras} \textbf{460}, 2834 (2016),
  \eprint{1605.05091}

\bibitem[{{Schmidt} et~al.(2016)}]{Schmidt2016}
{Schmidt}, T.~O.~B., et~al., \emph{{Direct Imaging discovery of a second planet
  candidate around the possibly transiting planet host CVSO 30}}, \emph{\aap}
  \textbf{593}, A75 (2016), \eprint{1605.05315}

\end{thebibliography}

\end{document}